\documentclass{ws-procs975x65}

\newcommand{\hi}{{\bf {H}}}

\newcommand{\dom}{{\bf {D}}}

\begin{document}

\title{Short-Distance Cutoffs in Curved Space\footnote{\uppercase{T}alk presented at
10th \uppercase{G}rossmann \uppercase{M}eeting on \uppercase{G}eneral
\uppercase{R}elativity, \uppercase{R}io de \uppercase{J}aneiro,
\uppercase{J}uly 2003.}}

\author{A. Kempf\footnote{\uppercase{T}his
work is supported by \uppercase{NSERC, CFI, OIT, PREA} and the
\uppercase{C}anada \uppercase{R}esearch \uppercase{C}hairs
\uppercase{P}rogram.}}

\address{Department of Applied Mathematics, University of Waterloo\\
Waterloo, Ontario N2L 3G1, Canada\\ akempf@uwaterloo.ca}

\maketitle

\abstracts{It is shown that space-time may possess the differentiability
properties of manifolds as well as the ultraviolet finiteness properties
of lattices. Namely, if a field's amplitudes are given on any sufficiently
dense set of discrete points this could already determine the field's
amplitudes at all other points of the manifold. The criterion for when
samples are sufficiently densely spaced could be that they are apart on
average not more than at a Planck distance. The underlying mathematics is
that of classes of functions that can be reconstructed completely from
discrete samples. The discipline is called sampling theory and is at the
heart of information theory. Sampling theory establishes the link between
continuous and discrete forms of information and is used in ubiquitous
applications from scientific data taking to digital audio.}

\section{Introduction}
General relativity and quantum theory, when considered together, are
indicating that the notion of distance loses operational meaning at the
Planck scale of about $10^{-35}m$ (assuming 3+1 dimensions). Namely, if
one tries to resolve a spatial structure with an uncertainty of less than
a Planck length, then the corresponding momentum uncertainty should
randomly curve and thereby significantly disturb the very region in space
that is meant to be resolved. To obtain a unified theory of general
relativity and quantum theory is difficult, however, not least because the
two theories are formulated in the very different languages of
differential geometry and functional analysis.

One of the problems in the effort to find a unifying theory of quantum
gravity is, therefore, to develop a mathematical framework which combines
differential geometry and functional analysis such as to give a precise
description of a notion of a shortest distance in nature. Candidate
theories may become testable when introduced to inflationary cosmology and
compared to CMB measurements, see \cite{cosm}.

There has been much debate about whether the unifying theory will describe
space-time as being discrete or continuous. It is tempting, also, to
speculate that a quantum gravity theory such as M theory, see \cite{pol},
or a foam theory, see \cite{foam}, once fully understood, might reveal the
structure of space-time as being in some sense in between discrete and
continuous, possibly such as to combine the the differentiability of
manifolds with the ultraviolet finiteness of lattices. This third
possibility seems to be ruled out, however: as G{\"o}del and Cohen proved,
no set can be explicitly constructed whose cardinality would be in between
discrete and continuous, see for example \cite{ak-paris}. Nevertheless,
there still is at least one possibility by which space-time might have a
mathematical description which combines the differentiability of manifolds
with the ultraviolet finiteness of lattices:

\section{Fields with a finite density of degrees of freedom}

Let us recall that physical theories are formulated not directly in terms
of points in space or in space-time but rather in terms of the functions
in space or in space-time. This suggests a whole new class of mathematical
models for a finite minimum length. Namely, fields in space-time could be
functions over a differentiable manifold as usual, while the class of
physical fields is such that if a field is sampled only at discrete points
then its amplitudes can already be reconstructed at {\it all} points in
the manifold - if the sampling points are spaced densely enough. The
maximum average sample spacing which allows one to reconstruct the
continuous field from discrete samples could be on the order of the Planck
scale, see \cite{ak-prl}.

Since any one of all sufficiently tightly spaced lattices would allow
reconstruction, no particular lattice would be preferred. It is because no
particular lattice is singled out that the symmetry properties of the
manifold can be preserved.

The physical theory could be written, equivalently, either as living on a
differentiable manifold, thereby displaying e.g. external symmetries, or
as living on any one of the sampling lattices of sufficiently small
average spacing, thereby displaying its ultraviolet finiteness. Physical
fields, while being continuous or even differentiable, would possess only
a finite density of degrees of freedom.

The mathematics of classes of functions which can be reconstructed from
discrete samples is well-known, namely as {\it sampling theory}, in the
information theory community, where it plays a central role in the theory
of sources and channels of continuous information as developed by Shannon,
see \cite{shannon}.
\section{Sampling theory}
The simplest example in sampling theory is the Shannon sampling theorem:
Choose a frequency $\omega_{max}$. Consider the class $B_{\omega_{max}}$
of continuous functions $f$ whose frequency content is limited to the
interval $(-\omega_{max},\omega_{max})$, i.e. for which:
%\begin{equation}
$ \tilde{f}(\omega)~=~\int_{-\infty}^{\infty} f(x) e^{-i\omega x} = 0
\mbox{~~~whenever ~~} \vert\omega\vert\ge\omega_{max}.$
%\end{equation}
If the amplitudes $f(x_n)$ of such a function are known at equidistantly
spaced discrete values $\{x_n\}$ whose spacing is $\pi/\omega_{max}$ or
smaller, then the function's  amplitudes $f(x)$ can be reconstructed for
all $x$. Shannon's reconstruction formula reads:
\begin{equation}
f(x)~ =~ \sum_{n=-\infty}^\infty ~f(x_n) ~
\frac{\sin[(x-x_n)\omega_{max}]}{(x-x_n)\omega_{max}}
\end{equation}
The theorem is in ubiquitous use in digital audio and video as well as in
scientific data taking. Sampling theory, see \cite{ferreira}, studies
generalizations of the theorem for various different classes of functions,
for non-equidistant sampling, for multi-variable functions and it
investigates the effect of noise, which could be quantum fluctuations in
our case. As was shown in \cite{ak-prl}, generalized sampling theorems
automatically arise from stringy uncertainty relations, namely whenever
there is a finite minimum position uncertainty $\Delta x_{min}$, as e.g.
in uncertainty relations of the type: $\Delta x \Delta p \ge
\frac{\hbar}{2}(1 + \beta (\Delta p)^2 +...)$, see \cite{ucr}. A few
technical remarks: the underlying mathematics is that of symmetric non
self-adjoint operators. Through a theorem of Naimark, unsharp variables of
POVM type arise as special cases.

Due to the origin of sampling theory in engineering, sampling theory for
generic (pseudo-) Riemannian manifolds has been undeveloped so far. Here,
I suggest a general approach to this problem.

\section{Information Theory on Curved Space}
Let us consider as a natural (because covariant) analogue of the bandwidth
restriction of the Shannon sampling theorem in curved space the presence
of a cutoff on the spectrum of the Laplace operator $-\Delta$ on a
Riemannian manifold ${\bf {M}}$. (On a pseudo-Riemannian or a spin
manifold, we can consider the d'Alembert or the Dirac operator, to obtain
a covariant form of sampling theory. This is pursued in \cite{ak-covsam}.)

We start with the usual Hilbert space $\hi$ of square integrable scalar
functions over the manifold, and we consider the dense domain
$\dom\subset\hi$ on which the Laplacian is essentially self-adjoint. Using
physicists' sloppy but convenient terminology we will speak of all points
of the spectrum as eigenvalues, $\lambda$, with corresponding
``eigenvectors" $\vert \lambda)$. Since we are mostly interested in the
case of noncompact manifolds, whose spectrum will not be discrete, some
more care will be needed, of course. For Hilbert space vectors we use the
notation $\vert \psi)$, in analogy to Dirac's bra-ket notation, only with
round brackets.

Let us define $P$ as the projector onto the subspace spanned by the
eigen\-spaces of the Laplacian with eigenvalues smaller than some fixed
maximum value $\lambda_{max}$. (For the d'Alembertian and for the Dirac
operator, let $\lambda_{max}$ bound the absolute values of the
eigenvalues.)

We consider now the possibility that in nature all physical fields are
contained within the subspace $\dom_s= P.\dom$, where $\lambda_{max}$
might be on the order of $1/l^2_{Planck}$. In fact, through this spectral
cutoff, each function in $\dom_s$ acquires the sampling property: if its
amplitude is known on a sufficiently dense set of points of the manifold,
then it can be reconstructed everywhere. Thus, through such a spectral
cutoff a sampling theorem for physical fields arises naturally. To see
this, assume for simplicity that one chart covers the $N$-dimensional
manifold. Consider the coordinates $\hat{x}_j$, for $j=1,...,N$ as
operators that map scalar functions to scalar functions: $\hat{x_j}:
\phi(x) \rightarrow x_j \phi(x)$. On their domain within the original
Hilbert space $\hi$, these operators are essentially self-adjoint, with an
``Hilbert basis" of non-normalizable joint eigenvectors $\{\vert x)\}$. We
can write scalar functions as $\phi(x)=( x\vert \phi)$, i.e. scalar
functions are the coefficients of the abstract Hilbert space vector $\vert
\phi)\in \hi$ in the basis of the vectors $\{\vert x )\}$. The continuum
normalization of the $\vert x)$ is with respect to the measure provided by
the metric. On the domain of physical fields, $\dom_s$, the multiplication
operators $\hat{x}_j$ are merely symmetric but not self-adjoint. The
projections $P\vert x)$ of the eigenvectors $\vert x )$ onto the physical
subspace $\dom_s$ are in general no longer orthogonal. Correspondingly,
the uncertainty relations of the first quantization must be modified, see
\cite{ak-old}.

Consider now a physical field, i.e. a vector $\vert \phi)\in \dom_s$,
which reads as a function: $\phi(x)=(x\vert\phi)$. Assume that only at the
discrete points $\{x_n\}$ the field's amplitudes $\phi(x_n)=\langle
x_n\vert \phi\rangle$ are known. Then, if the discrete sampling points
$\{x_n\}$ are sufficiently dense, they fully determine the Hilbert space
vector $\vert \phi\rangle$, and therefore $\phi(x)$ everywhere. To be
precise, we assume the amplitudes
\begin{equation}
\phi(x_n)=(x_n\vert\phi)=\sum_{\vert\lambda\vert<\lambda_{max}}
\!\!\!\!\!\!\!\!\!\!\!\!\!\!\int ~(x_n\vert \lambda)(\lambda \vert \phi)
\end{equation}
 to be known. We use the sum and
integral notation because $\{\lambda\}$ may be discrete and or continuous
(the manifold ${\bf {M}}$ may or may not be compact). Define
$K_{n\lambda}= (x_n\vert\lambda)$. The set of sampling points $\{x_n\}$ is
dense enough for reconstruction iff $K$ is invertible, because then:
$(\lambda\vert \phi)=\sum_n K^{-1}_{\lambda,n} \phi(x_n)$. We obtain the
reconstruction formula:
\begin{equation}
\phi(x) = \sum_n \left(\sum_{\vert\lambda\vert<\lambda_{max}}
\!\!\!\!\!\!\!\!\!\!\!\!\!\!\int ~ (x\vert\lambda) K^{-1}_{\lambda
n}\right) ~\phi(x_n)
\end{equation}
It should be interesting to investigate the stability of the
reconstruction of samples along the lines of the work by Landau
\cite{landau}. The reconstruction stability is usually of importance due
to noise. Here the question arises to which extent quantum fluctuations
may play the role of noise. Also, following Shannon and Landau, it appears
natural to define the density of degrees of freedom through the number of
dimensions of the space of functions in $\dom_s$ with essential support in
a given volume. Clearly, we recover conventional Shannon sampling as a
special case. The Shannon case has been applied to inflationary cosmology
in \cite{cosm} and it should be very interesting to apply to cosmology
also the more general approach presented here. We note that higher than
second powers of the fields (second powers occur as scalar products in the
Hilbert space of fields) are now nontrivial to enter into a quantum field
theoretical action: To this end, the multiple product of fields needs to
be defined such as to yield a result within the cut-off Hilbert space. In
this context it should be interesting also to reconsider the mechanism of
Sakharov's induced gravity, see \cite{sakharov}.

Our approach to sampling on curved space significantly simplifies in the
case of compact manifolds, where the spectrum of the Laplacian is discrete
and the cut off Hilbert space $\hi_s$ is finite dimensional. Intuitively,
it is clear that knowledge of a function at as many points as is the
dimension of the cutoff Hilbert space generically allows one to
reconstruct the function everywhere. Results of spectral geometry, see
e.g. \cite{specs1,specs2,specs3,specs4} are likely to be useful also for
the approach described here.

\section{Vacuum energy}

Vacuum energy is of great current cosmological interest and it will be
interesting to use the covariant cutoff tools here to investigate this
problem. Regarding the vacuum energy, I would like to close with a
speculative thought: In the Casimir effect, the system's energy can be
lowered not only by moving the plates closer but instead also by making
plates which are at a fixed distance into better conductors, as this
increases the strength and frequency range of their mode-expelling
property. It should be interesting to investigate if the copper oxide
``plates" in High-Tc superconductors can be viewed as being driven towards
superconductivity by the reduction in energy through the Casimir effect.

\end{document}